# The Isgur-Wise Function from the Lattice

Laurent Lellouch (UKQCD Collaboration)

*Physics Department, The University*
*Southampton, SO9 5NH, United Kingdom*

## ABSTRACT

We calculate the Isgur-Wise function by measuring the elastic scattering amplitude of a $D$ meson in the quenched approximation on a $24^3 \times 48$ lattice at $\beta = 6.2$, using an $\mathcal{O}(a)$-improved fermion action. We use this result, in conjunction with heavy-quark symmetry, to extract $|V_{cb}|$ from the experimentally measured $\bar{B} \to D^* l \bar{\nu}$ differential decay width.

Heavy-quark symmetry enables all the non-perturbative, strong-interaction physics for semi-leptonic $\bar{B} \to D, D^*$ decays to be parametrized in terms of a single universal function, $\xi(\omega)$, of $\omega \equiv v \cdot v'$, where $v$ and $v'$ are the four-velocities of the $\bar{B}$ and $D$ mesons respectively[1, 2]. $\xi(\omega)$ is known as the Isgur-Wise function and is normalized at zero recoil: $\xi(1) = 1$[2]. With $\xi(\omega)$, one can extract the Cabibbo-Kobayashi-Maskawa matrix element $V_{cb}$ from experimental measurements of the rate for $\bar{B} \to D^* l \bar{\nu}$ decays[3]. We report here on a lattice QCD calculation of the Isgur-Wise function and on the corresponding determination of $V_{cb}$. We also perform a qualitative test of the flavor component of heavy-quark symmetry.

To obtain the Isgur-Wise function, we evaluate the elastic scattering matrix element $\langle D(p')|\bar{c}\gamma^\mu c|D(p)\rangle$ on the mass shell[4]. Because the electromagnetic current $\bar{c}\gamma^\mu c$ is conserved, this matrix element can be parametrized in terms of a single form factor:

$$\langle D(p')|\bar{c}\gamma^\mu c|D(p)\rangle = m_D(v+v')^\mu h^{el}(\omega) , \quad (1)$$

where $p^{(\prime)} = m_D v^{(\prime)}$ and $\omega = v \cdot v'$ is the four-velocity recoil. In the limit of exact heavy-quark symmetry this form factor is simply $\xi(\omega)$.

There are two sources of corrections to this simple result:

$$h^{el}(\omega) = \left[1 + \beta^{el}(\omega) + \gamma^{el}(\omega)\right] \xi(\omega) . \quad (2)$$

The first correction, $\beta^{el}(\omega)$, stems from perturbative QCD corrections to the heavy-quark current. We obtain this correction from Neubert's short distance expansion of heavy-quark currents[5]. The second correction, $\gamma^{el}(\omega)$, is due to higher-dimension operators with coefficients proportional to inverse powers of the charm quark mass. $\gamma^{el}(\omega)$ is difficult to quantify because it involves the light degrees of freedom and is therefore non-perturbative. Luke's theorem[6], however, guarantees that there is no $\mathcal{O}(\Lambda_{QCD}/(2m_c))$ correction to $h^{el}(\omega)$ at zero recoil. Moreover, model estimates of this correction appears to remain well below 3% over the range of experimentally accessible recoils[7]. That this correction is small is corroborated by the fact that we see no differences in the ratio $h^{el}(\omega)/(1+\beta^{el}(\omega))$ computed for two values of the heavy-quark mass. Thus, we will neglect $\gamma^{el}(\omega)$ in extracting the Isgur-Wise function from $h^{el}(\omega)$. As defined in Eq. (2), $\xi(\omega)$ is renormalization-group invariant and normalized to one at $\omega = 1$[5].

We work in the quenched approximation on a $24^3 \times 48$ lattice at $\beta = 6.2$, which corresponds to an inverse lattice spacing $a^{-1} = 2.73(5)$ GeV, as determined from the string tension[8]. Our calculation is performed on sixty $SU(3)$ gauge field configurations (for details see Ref. [8]). The mesons are composed of a propagating heavy quark with a mass around that of the charm quark, and a light antiquark with a mass around that of the strange quark. To reduce discretization errors, the quark propagators are calculated using an $\mathcal{O}(a)$-improved action[9]. This improvement is particularly important here since we are studying the propagation of quarks whose bare masses are around one half the inverse lattice spacing. Our statistical errors are calculated according to the bootstrap procedure described in Ref. [8].

To obtain the matrix element $\langle D(\mathbf{p}')|\bar{c}\gamma^\mu c|D(\mathbf{p})\rangle$, we calculate the ratio of three-point correlators,

$$A^\mu(t_x) \equiv \frac{\sum_{\mathbf{x},\mathbf{y}} e^{-i(\mathbf{q}\cdot\mathbf{x}+\mathbf{p}'\cdot\mathbf{y})}\langle J(y) V^\mu(x) J^\dagger(0)\rangle}{\sum_{\mathbf{x},\mathbf{y}} \langle J(y) V^0(x) J^\dagger(0)\rangle} , \quad (3)$$

where $J$ is a spatially-extended interpolating field for the $D$ meson[10], $V^\mu$ is the $\mathcal{O}(a)$-improved version of the vector current $\bar{c}\gamma^\mu c$[11] and $\mathbf{p}=\mathbf{p}'+\mathbf{q}$. To evaluate these correlators, we use the standard source method[12]. We choose $t_y = 24$ and symmetrize the correlators about that point using Euclidean time reversal. We evaluate $A^\mu$ for three values of the light-quark mass ($\kappa_l = 0.14144, 0.14226, 0.14262$) which straddle the strange quark mass (given by $\kappa_s = 0.1419(1)$[13]); two values of the heavy-quark mass ($\kappa_h = 0.121, 0.129$) around that of the charm quark (given by $\kappa_c \simeq 0.129$[14]); and several values of the initial and final $D$-meson momenta, all less than $(\pi/12a)\sqrt{2}$.

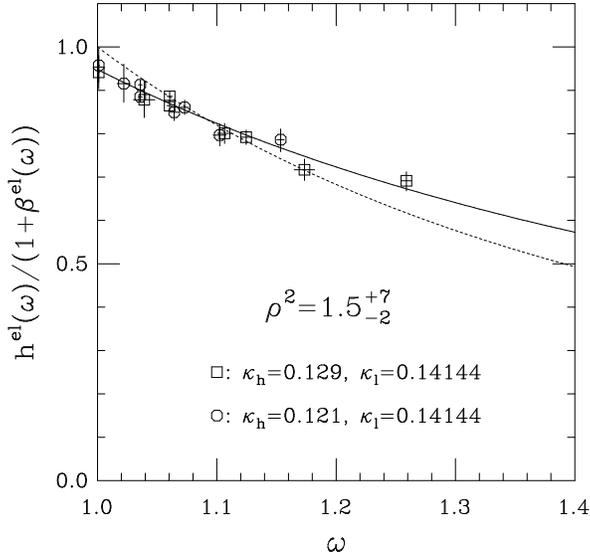

Figure 1: $h^{el}(\omega)/(1+\beta^{el}(\omega))$ vs $\omega$.

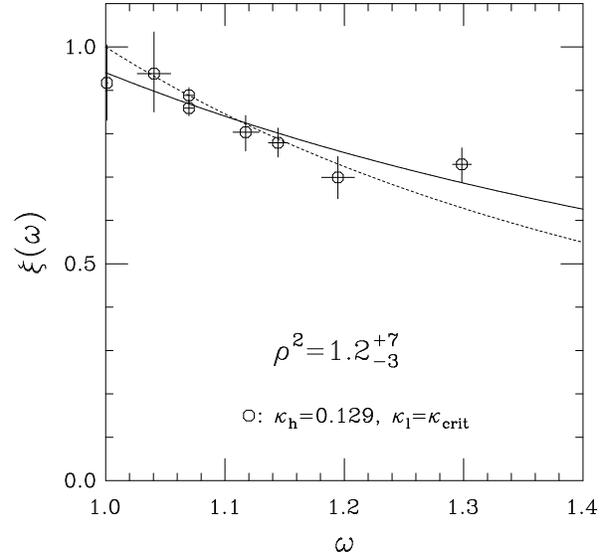

Figure 2: $\xi(\omega)$ vs $\omega$.

In the limit $t_x, t_y - t_x \to \infty$, we fit to

$$R(t_x) = A^0(t_x) e^{(E-E')t_x + (E'-m_D)t_y} \longrightarrow$$
$$\frac{m_D}{2EE'} \frac{Z(\mathbf{p}^2)Z(\mathbf{p'}^2)}{Z^2(0)} \langle D(\mathbf{p'})|\bar{c}\gamma^0 c(0)|D(\mathbf{p})\rangle, \qquad (4)$$

where $E$ ($E'$) is the energy of the initial (final) $D$ meson and $Z(\mathbf{p}^2) \equiv \langle 0|J(0)|D(\mathbf{p})\rangle$. We see a plateau in $R$ about $t_x = 12$, and fit $R$ to a constant for $t_x = 11, 12, 13$. Multiplying this constant by suitable wave-function and energy factors, obtained from fits to two-point functions, we get $\langle D(\mathbf{p'})|\bar{c}\gamma^0 c|D(\mathbf{p})\rangle$.

The ratio $h^{el}(\omega)/(1 + \beta^{el}(\omega))$ that we obtain for $\kappa_l = 0.14144$, the heaviest of our light quarks, is plotted in Fig. 1. Data for both values of the heavy-quark mass, $m_h$, are shown. The squares correspond to $m_{h1} = 1.4\,\mathrm{GeV}$ ($k_h = 0.129$) and the octogons to $m_{h2} = 1.9$ GeV ($k_h = 0.121$)[*]. The two sets of data agree very well. It is difficult, however, to use this information to quantify the $1/m_c$ correction since our range of heavy-quark masses is small. Only $(m_{h2} - m_{h1})/m_{h2} \simeq 1/4$ of the full $1/m_c$ correction can be resolved. The fact that we see no sign of this correction indicates that it must be smaller than four times a few percent. In Fig. 1, the solid curve is a two-parameter fit to $s\xi_\rho(\omega)$, where $\xi_\rho(\omega)$ is Stech's relativistic-oscillator parametrization[7]:

$$\xi_\rho(\omega) = \frac{2}{\omega+1} \exp\left(-(2\rho^2-1)\frac{\omega-1}{\omega+1}\right) \qquad (5)$$

and $\rho^2 = -\xi'_\rho(1)$. The parameter $s$ was added to absorb uncertainties in the overall normalization of our data.

---

[*]$m_{h1(h2)}$ is the spin averaged mass of pseudoscalar and vector mesons composed of a heavy-quark, $h_1(h_2)$, and a massless antiquark[14], less 0.5 GeV (p. 79 of Ref. [15])

We find $\rho^2 = 1.5^{+2}_{-2}$ and $s = 0.95^{+1}_{-1}$ with a $\chi^2/\mathrm{dof} = 0.8$. Other parametrizations for $\xi(\omega)$ give very similar results.

If we fit our data to $\xi_\rho(\omega)$ instead of $s\xi_\rho(\omega)$ (an equally valid procedure, in principle, for determining $\rho^2$), we find $\rho^2 = 2.1^{+1}_{-1}$ with a $\chi^2/\mathrm{dof} = 2.9$ (dotted curve in Fig. 1). To accommodate the spread in values for $\rho^2$ given by our two procedures, we assign errors to $\rho^2$ which encompass all values consistent with both procedures. These errors include systematic uncertainties, but only to the extent that the deviation of $s$ from 1, in our first fit, is an indication of systematic errors. The central value we choose for $\rho^2$ is the one given by our first fit since this fit is designed to absorb uncertainties in the overall normalization of our data. Thus, for $\kappa_l = 0.14144$ we quote $\rho^2 = 1.5^{+7}_{-2}$.

In Fig. 2 we plot the results obtained from a covariant and linear extrapolation of our data for $h^{el}(\omega)/(1+\beta^{el}(\omega))$ for the three values of the light-quark mass to $\kappa_l = \kappa_{crit}$. These results correspond to a meson composed of a charm quark ($\kappa_h = 0.129$) and a massless antiquark. They are our Isgur-Wise function. With a two parameter fit to $s\xi_\rho(\omega)$, we get $\rho^2 = 1.2^{+3}_{-3}$ and $s = 0.94^{+2}_{-2}$ with a $\chi^2/\mathrm{dof} = 0.9$ (solid line in Fig. 2). Forcing $s$ to be 1, we find $\rho^2 = 1.7^{+2}_{-2}$ with a $\chi^2/\mathrm{dof} = 1.6$ (dotted line in Fig. 2). Using the same procedure as the one used to determine the errors and central value for $\rho^2$ when $\kappa_l = 0.14144$, we obtain, as our best estimate for $\rho^2$, when $\kappa_l = \kappa_{crit}$:

$$\rho^2 = 1.2^{+7}_{-2}. \qquad (6)$$

Dependence of $\xi(\omega)$ on light-quark mass is mild. Fur-thermore, our result for $\rho^2$ agrees with most other determinations of this parameter[3, 16], apart from the sum-rule result of Ref. [17] which lies below our error bars. In particular, our result for $\rho^2$ agrees with the very recent lattice result of Ref. [18] obtained with Wilson fermions, although the details and systematics of the two calculations are different.

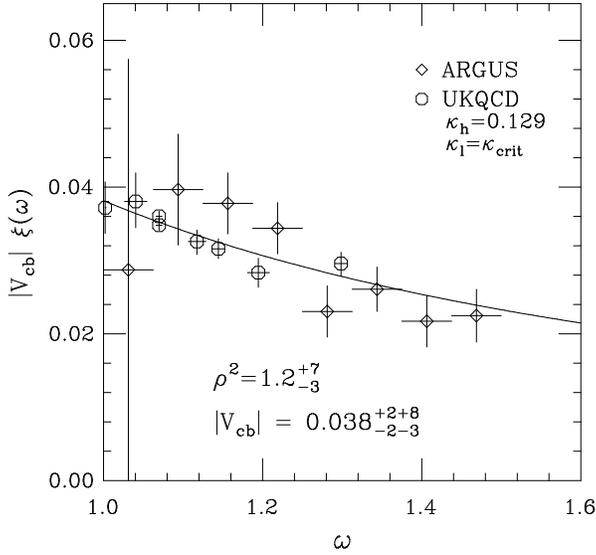

Figure 3: $|V_{cb}|\xi(\omega)$ vs $\omega$. The solid line is $|V_{cb}|\xi_\rho(\omega)$ with $\rho^2 = 1.2$.

We can now obtain $|V_{cb}|$. Heavy-quark symmetry enables one to write the differential decay rate for $\bar{B} \to D^*l\bar{\nu}$ decays in terms of the Isgur-Wise function:

$$\frac{d\Gamma}{d\omega} \propto |V_{cb}|^2 \xi(\omega)^2 \ , \qquad (7)$$

where the proportionality factors are described in Ref. [3]. Thus, experimental data for $d\Gamma/d\omega$ yield a measurement of $|V_{cb}|\xi(\omega)$ versus $\omega$. To get $|V_{cb}|$, we fit $|V_{cb}|\xi_\rho(\omega)$ to these data with $\xi_\rho(\omega)$ given by Eq. (5) and $\rho^2$ fixed by our lattice calculation to the value given in Eq. (6). In Fig. 3 we show this least-$\chi^2$-fit for new ARGUS data[19]. The resulting value for $|V_{cb}|$ is

$$|V_{cb}|\sqrt{\tfrac{\tau_B}{1.48\mathrm{p}s}} = 0.038^{+2\,+8}_{-2\,-3} \ , \qquad (8)$$

with a $\chi^2/\mathrm{d}of = 1.1$. The same fit to the weighted average of older CLEO and ARGUS data[20] gives $|V_{cb}|\sqrt{\tau_B/1.48\mathrm{p}s} = 0.036^{+2\,+8}_{-2\,-3}$ with a $\chi^2/\mathrm{d}of = 0.6$. In both cases, the first set of errors is due to experimental uncertainties, while the second is due to the uncertainty in our lattice determination of $\rho^2$. The B-meson lifetime used above is the central value of the lifetime, $\tau_B^0 = 1.48(10)\mathrm{p}s$, quoted in Ref. [21].